\documentclass[10pt,twocolumn]{article}

\usepackage[utf8]{inputenc}
\usepackage{textgreek}
\usepackage{amsmath}
\usepackage{amssymb}
\usepackage{amsthm}
\usepackage{titlesec}
\usepackage{cite}
\usepackage{color}
\usepackage{units}
\usepackage{booktabs}
\usepackage{graphicx}
\usepackage[colorlinks=false]{hyperref}
\usepackage[format=plain,labelfont=it]{caption}
\usepackage[left=1.5cm,right=1.5cm,top=2cm,bottom=2cm]{geometry}
\usepackage{threeparttable}
\usepackage{multirow} 
\usepackage{adjustbox}
\usepackage{tikz}
\usetikzlibrary{arrows}
\usetikzlibrary{shapes}
\usetikzlibrary{decorations.pathmorphing}
\usetikzlibrary{decorations.pathreplacing}
\usetikzlibrary{decorations.shapes}
\usetikzlibrary{decorations.text}
\usetikzlibrary{positioning, shapes}
\usetikzlibrary{calc}
\tikzset{
block/.style = {draw, fill=white, rectangle, minimum height=3em, minimum width=4.5em},
tmp/.style  = {coordinate}, 
sum/.style= {draw, fill=white, circle, node distance=2cm},
input/.style = {coordinate},
output/.style= {coordinate}
pinstyle/.style = {pin edge={to-,thick,black}}
}
\usepackage[labelformat=empty]{subfig}

\titleformat{\section}{\centering\normalfont\scshape}{\Roman{section}.}{5pt}{}
\titleformat{\subsection}{\normalfont\it}{\Alph{subsection}.}{5pt}{}
\titleformat{\subsubsection}{\normalfont\it}{\hspace{4mm}\arabic{subsubsection})}{5pt}{}

\newcommand\infoFootnote[1]{%
  \begingroup
  \renewcommand\thefootnote{}\footnote{#1}%
  \addtocounter{footnote}{-1}%
  \endgroup}

\theoremstyle{remark}

\newcommand{\R}{\mathbb{R}} 
\newcommand{\N}{\mathbb{N}}

\newcommand{\Z}{\mathbb{Z}}

\newcommand{\ab}{\boldsymbol{a}}
\newcommand{\db}{\boldsymbol{d}}

\newcommand{\bb}{\boldsymbol{b}}

\newcommand{\hb}{\boldsymbol{h}}

\newcommand{\thetab}{\boldsymbol{\theta}}

\newcommand{\oneb}{\boldsymbol{1}}

\newcommand{\modu}{\,\mathrm{mod}\,}

\newcommand{\xt}{\mathtt{x}}
\newcommand{\yt}{\mathtt{y}}
\newcommand{\Enc}{\mathtt{Enc}}
\newcommand{\Dec}{\mathtt{Dec}}

\title{\LARGE \bf
Encrypted extremum seeking for\\ privacy-preserving PID tuning as-a-Service
}
\date{}

\author{Nils Schl\"uter, Matthias Neuhaus, and Moritz Schulze Darup%
\thanks{N.~Schl\"uter, M. Neuhaus, and M.~Schulze Darup are with the Automatic Control Group, Department of Mechanical Engineering, TU Dortmund University, Germany.
        {E-mails:  \{nils.schlueter,
        matthias.neuhaus,
        moritz.schulzedarup\}@tu-dortmund.de}.}
}

\begin{document}

\maketitle
    \thispagestyle{empty}
\pagestyle{empty}

\textbf{\textit{Abstract}.} {\bf 
Wireless communication offers many benefits for control such as substantially reduced deployment costs, higher flexibility, as well as easier data access. It is thus not surprising that smart and wireless sensors and actuators are increasingly used in industry.
With these enhanced possibilities, exciting new technologies such as Control-as-a-Service arise, where (for example) controller design or tuning based on input-output-data can be outsourced to a cloud or mobile device. This implies, however, that sensitive plant information may become available to service providers or, possibly, attackers.

Against this background, we focus on privacy-preserving optimal PID tuning as-a-Service here. In particular, we combine homomorphic encryption with extremum seeking in order to provide a purely data-driven and confidential tuning algorithm. The encrypted realization requires several adaptions of established extremum seekers. These encompass relative parameter updates, stochastic gradient approximations, and a normalized objective function. As a result, and as illustrated by various numerical examples, the proposed encrypted extremum seeker is able to tune PID controllers for a wide variety of plants without being too conservative.
}
\infoFootnote{\hspace{-1.5mm}$^\ast$This paper is a \textbf{preprint} of a contribution to the 20th European Control Conference 2022.}

\section{Introduction}
\label{sec:intro}
Wireless communication is of growing interest in the automation and process industry due to, e.g., cost reduction, high flexibility, and easy data access. It is often viewed as a key enabler of industry 4.0 and plays a vital role in fields like mobile robotics, building automation, and intelligent transportation systems. Clearly, wireless communication also effects control loops, where it becomes possible to make extensive use of input-output-data provided by smart sensors and actuators. This also enables Control-as-a-Service (CaaS), where plant data is processed externally on a cloud or a mobile device in order to realize remote control or remote controller design and tuning. CaaS offers many advantages compared to classical in-house control engineering. For instance, it allows reducing the necessity of expert knowledge and the effort for time-consuming design processes. Moreover, elaborated design methods can be used, yielding high-performance controllers. However, CaaS is not yet widely used since it involves the exchange of and computations on highly sensitive plant data. 

In principle, privacy-preserving CaaS can be realized with the help of modern cryptosystems such as homomorphic encryption (HE) (see, e.g., \cite{Paillier1999,Gentry2010,miccianciofhew,Cheon2017_CKKS}) which enable (simple) computations on encrypted data. Utilizing HE for encrypted control is, however, non-trivial and usually requires tailored controller reformulations (see \cite{SchulzeDarup2021_CSM} for an overview). Still, various control schemes have already been successfully (re)implemented in an encrypted fashion (see, e.g., \cite{SchulzeDarup2018_LCSS,Alexandru2018_CDC,SchlueterDarupECC2021}). Nevertheless, a more specific analysis of existing realizations shows that encrypted controller design or tuning has rarely been considered. One exception is the recently proposed HE-based implementation of a data-driven LQR from~\cite{alexandru2020deepc} which  builds on the scheme in~\cite{doerfler2019deepc}. In order to strengthen this promising direction of encrypted data-driven control, we aim for secure cloud-based tuning of PID controllers in this paper.

Our focus on PID tuning as-a-Service is motivated by two key observations. First, PID controllers are still widely used in industry. In fact, according to~\cite{samad2017survey}, they are the control technology with the highest (perceived) impact. Second, despite their simple  structure, a proper PID tuning can be challenging and time-consuming. Consequently, many PID controllers are tuned by hand, by heuristics such as Ziegler and Nichols, or aren't tuned at all. This results (often) in a performance that is far from optimal. Against this background, our goal is to provide a cloud-based PID tuning that preserves privacy of the exchanged and processed plant data.
 
Now, there exist a vast amount of PID tuning methods with different scopes, and a competitive cloud-service should probably offer various methods to choose from. However, we aim for a proof of concept here and, hence, we will concentrate on a privacy-preserving implementation of one scheme which can be efficiently implemented with current HE techniques. In fact, model-based approaches such as pole placement, loop-shaping, or internal model control are hard to realize in an encrypted fashion if model identification is part of the cloud-service. At the same time, the lack of a model and the need for ad-hoc controller designs/improvements is what motivates to use CaaS. As a consequence, we focus on purely data-driven tuning methods at the cost of stability guarantees. Nonetheless, the following methods provide a handle on rapid improvements of a PID's control performance and have been extensively tested on a variety of closed-loops. Possible candidates are, for example, iterative feedback tuning, virtual reference feedback tuning, correlation-based tuning, or extremum seeking (ES) control~\cite{krstic2003real}. A comparison of these schemes reveals that especially the latter combines an encryption-friendly algorithm with an optimization based tuning approach. Thus, the remaining paper will focus on HE-based PID tuning via ES. 

\textbf{Road map.} In Section~\ref{sec:preliminaries} we introduce ES with the  focus on PID controller tuning and shed some light on HE. With this knowledge at hand, we tackle the reformulation of ES tailored for encryption in Section~\ref{sec:main}. Finally, we test our proposed method with numerical examples in Section~\ref{sec:benchmark} and end the paper with a conclusion and outlook in Section~\ref{sec:outlook}, respectively.

\textbf{Notation.} We denote the modulo operation $z \modu q:=z-q\lfloor z/q \rfloor$. In this context, $\lfloor\cdot \rfloor$, $\lceil\cdot \rceil$, and $\lfloor\cdot \rceil$ refer to the floor function, the ceiling function, and rounding to the nearest integer, respectively. Moreover, the Hadamard product, i.e., component-wise multiplication, of two vectors $\ab\in\R^{n}$ and $\bb\in\R^{n}$ is denoted by $\ab\circ\bb$ and the vector of ones in suitable dimension is $\oneb$.

\section{Preliminaries}
\label{sec:preliminaries}
We begin with a brief summary on how ES can be used for controller tuning. In general, ES is a model-free online optimization technique. As such, it is often applied if a model is unavailable or unreliable, but also if the optimization criterion is (slowly) changing with time. As we will shortly see, ES is a very simple algorithm that can be easily adapted for controller tuning. This adaptation will be the basis for our privacy-preserving PID tuning in Section~\ref{sec:main}. The corresponding cryptographic foundations will be laid in Section~\ref{subsec:homomorph}, where we collect some useful insights on HE.

\subsection{Extremum seeking control}
\label{subsec:extremumseeking}
ES is used to determine or maintain an extremum value $J(\thetab^\ast)$ of an objective function $J(\thetab):\R^{p}\rightarrow \R$ with parameters $\thetab\in\R^{p}$. This is achieved by perturbing $\thetab$ at iteration step $k\in \N$ according to
\begin{equation}
\label{eq:perturb}
        \thetab_{\db}(k):=\thetab(k)+\db_{\thetab}(k)
\end{equation}
with $\db_{\thetab}(k)$ being a user-defined perturbation. 
Clearly, this results in an ``exploration'' of $J(\thetab_{\db}(k))$ and, hence, allows approximating the gradient $\nabla_{\thetab} J(\thetab(k))$ (or even higher order derivatives) in the vicinity of $\thetab(k)$. Based on this information, the parameters are updated by, e.g., a gradient descent step of the form
\begin{equation}
    \label{eq:gradientstep}
        \thetab(k+1):=\thetab(k)+\Delta \thetab(k),
\end{equation}
where $\Delta \thetab(k):=-\alpha \nabla_{\thetab} J(\thetab(k))$ and $\alpha\in\R_{+}$ denotes the step width. By repeating these steps for suitable $\alpha$, the algorithm approaches a local minimum $J(\thetab^\ast)$.

\textbf{Controller tuning.} In classical setups, $J(\thetab)$ is often a cost function that characterizes an optimal operating point, e.g., for power tracking in photovoltaic systems or for maximizing the output of bioreactors~\cite{krstic2003real}. However, ES is also useful for PID  tuning~\cite{killingsworth2006pid}. There, the integrated square error 
\begin{equation}
\label{eq:costintegral}
    J(\thetab):=\int e^2(t,\thetab)\mathrm{d}t
\end{equation}
is minimized over a user-defined horizon, where $e(t,\thetab):=r(t)-y(t,\thetab)$ is the control error, i.e., the difference between the reference $r(t)$ and the plant output $y(t,\thetab)$ at time $t\in \R$, and where $\thetab=( K_p \,\,\,\, K_i \,\,\,\, K_d)^\top$ reflects the tuning parameters of the PID 
\begin{equation}
        \label{eq:PID}
        C(s,\thetab):=K_p+K_i\frac{1}{s}+K_d s.
\end{equation}
Remarkably, the extremum seeker has only access to closed-loop information in terms of~\eqref{eq:costintegral}. In fact, additional knowledge about the plant itself is not assumed.

Now, an important difference between classical ES control and ES-based controller tuning is that perturbations $\thetab_{\db}(k)$ of the controller parameters will not (or only insignificantly) excite the closed-loop in steady state. Hence, additional reference steps of the form
\begin{equation}
    \label{eq:stepinput}
r(t)=\begin{cases}
\hat{r} \text{ if } t\geq t_0\\
0 \text{ else}
\end{cases}
\end{equation}
are considered in~\cite{killingsworth2006pid} throughout the tuning process. However, other reference signals are possible. Yet, it should be noted that $\nabla_{\thetab}J(\thetab)$ depends on $r(t)$ by means of~\eqref{eq:costintegral}.

\textbf{Tuning setup.} In principle, we have now collected all crucial components for ES-based PID tuning. However, it remains to detail the choice of the perturbations $\db_{\thetab}(k)$ and the approximation of the gradient $\nabla_{\thetab}J(\thetab(k))$. The $i\in\{1,\dots,p\}$ perturbation signals $\left(\db_{\thetab}\right)_i$ are in general periodic functions with an amplitude of $\gamma_i\in\R$, zero mean, and non-zero squared average (power), e.g., cosine or square waves.
With $\db_{\thetab}$ at hand, the approximation of $\nabla_{\thetab}J(\thetab(k))$ works as follows. First, the controller parameters are updated by $\thetab_{\db}(k)$ and $r(t)$ is then used to excite the closed-loop. This allows to measure $e(t,\thetab)$ over some time frame and to obtain $J(\thetab_{\db}(k))$ by means of~\eqref{eq:costintegral}. From here on, depending on the choices for $\db_{\thetab}$, different approaches are possible. For instance, if $\db_{\thetab}$ are cosine wave perturbations, $\nabla_{\thetab}J(\thetab(k))$ can be estimated by a high-pass filter and a demodulation signal. This, more classic approach, is used in~\cite{killingsworth2006pid}. Square wave perturbations, on the other hand, enable numerical differentiation~\cite{nesic2015stochapprox}. Finally, $\alpha$ and $\gamma_i$ control the convergence speed and exploration, respectively. Thus, they must be chosen with care such that the closed-loop does not become unstable during~\eqref{eq:perturb} or~\eqref{eq:gradientstep} and the convergence speed is acceptable. For different closed-loop dynamics, these often have to be adapted manually. A convergent controller tuning will, however, improve the controller's performance by $J(\thetab)-J(\thetab^\ast)$.

\subsection{Homomorphic encryption for confidential computations}
\label{subsec:homomorph}

In a nutshell, HE allows carrying out (simple) mathematical operations on encrypted data (see~\cite{kim2020comprehensive} for an introduction). Over the previous decade, various schemes have been proposed with different functionalities (see, e.g,~\cite{Paillier1999,Gentry2010,miccianciofhew,Cheon2017_CKKS}). Commonly supported homomorphisms are encrypted additions and (or) multiplications. More precisely, let $\xt_1$ and $\xt_2$ be two arbitrary numbers in the cryptosystem's plaintext space and let us denote the encryption by ``$\Enc$'' and the decryption by ``$\Dec$''. Then, the homomorphisms ``$\otimes$'' and ``$\oplus$'' allow computing
\begin{subequations}
\label{eq:homomorphism}
\begin{align}
\label{eq:homtimes}
\xt_1 \, \xt_2 &= \Dec \left( \Enc(\xt_1) \otimes \Enc(\xt_2) \right), \\
\label{eq:homplus}
 \xt_1 + \xt_2 &= \Dec \left( \Enc(\xt_1) \oplus \Enc(\xt_2) \right).
\end{align}
\end{subequations}
Some further operations can  easily be derived from these fundamental ones. For instance, we will use ``$\ominus$'' for encrypted subtractions and ``$\boxtimes$'' for encrypted multiplications with one public factor $\xt_1$ or $\xt_2$. Obviously, with these homomorphisms at hand, polynomial expressions can be evaluated easily.
However, non-polynomial expressions are computationally expensive and typically require ``tricks''.

\textbf{Plaintext space and mapping.} The plaintext space of HE cryptosystems is usually some finite set with (for example) the canonical representatives
$$
    \Z_q:=\left\{0,\dots,q-1\right\},
$$
where $q\in \N_{>1}$ is the ciphertext modulus, i.e., the number of elements in the set $\Z_q$. Hence, before one can use HE, all quantities must be mapped to or encoded in $\Z_q$. An easy way to do this is as follows. For some scalar $x\in\R$, we compute
 \begin{align}
 \label{eq:map}
     \xt:=\lfloor c x \rceil \modu q,
 \end{align}
where $c\in\R_{\geq 1}$ is a scaling factor. In order to reconstruct $x$, one can use
\begin{equation}
\label{eq:reconstruction}
   x\approx \mu(\xt)/c \;\text{ with }\; \mu(\xt):=\begin{cases}
 \xt-q &\text{if } \xt\geq q/2 \\
 \xt &\text{else},
\end{cases}  
\end{equation}
where $|x-\mu(\xt)/c|\leq 1/(2c)$ if $\lfloor c x \rceil \in\Z_q$. For multidimensional quantities, i.e., vectors or matrices, \eqref{eq:map} and~\eqref{eq:reconstruction} can be used component-wise. Now, it is possible to investigate ciphertext additions and multiplications by means of plaintext additions and multiplications over $\Z_q$ due to~\eqref{eq:homomorphism}. Then, note that, e.g., $\mu\left(\xt^2+\yt\right)/c^2$ is meaningful for $\xt$ as in~\eqref{eq:map} only if $\yt:=\lfloor c^2 y\rceil \modu q$. In other words, summands should share the same scaling and multiplications increase the scaling. Due to the latter, $q$ and $c$ must be chosen carefully such that overflows, i.e., $\left(\xt^2+\yt\right)\notin\Z_q$, are avoided while maintaining sufficient accuracy. Furthermore, for arbitrarily many multiplications with $c>1$ and finite $q$, the plaintext will (almost certainly) overflow $\Z_q$, which, e.g., poses a problem for the operating time of a controller~\cite{SchlueterDarupECC2021}.

\textbf{LWE-based cryptosystems.} 
Cryptosystems providing~\eqref{eq:homomorphism} can, e.g., be constructed based on learning with errors (LWE) \cite{Regev2010learning}. These cryptosystems combined with a regular execution of a technique called ``bootstrapping'' (see \cite{Gentry2010}) results in a fully homomorphic cryptosystem and resolves the aforementioned problem regarding operating time. Fully HE (theoretically) allows computing any function in terms of boolean or arithmetic circuits. However, bootstrapping is computationally costly (many seconds to minutes up to now in the arithmetic case) such that fully HE is not yet useful for our application. Opposed to fully HE, a leveled HE scheme waives on bootstrapping [but also provides~\eqref{eq:homomorphism}], which results in performance advantages.
Consequently, only a finite number of operations\footnote{Additionally to the scale factor growth during multiplications, LWE-based cryptosystems inject a small error in the ciphertexts which grows during operations such that additions and multiplications can only be evaluated finitely often.} can be supported, i.e., for a predefined $c$, one must \mbox{a priori} select a large enough $q$ for the arithmetic circuit to be encrypted. 
Finally, we point out that among the leveled HE schemes, \cite{Cheon2017_CKKS} stands out because multiple optimizations such as constructions over polynomial rings with tailored plaintext encodings, residue number systems, or ciphertext rescaling make it somewhat practical. A detailed explanation of the cryptosystem or these methods is, however, out of this scope of this paper. 

\section{Robust encrypted extremum seeking}
\label{sec:main}

\subsection{Challenges for encrypted extremum seeking}
\label{subsec:challenges}
Several challenges have to be overcome in order to realize encrypted ES-based PID tuning as-a-Service. First (and probably most important), we need to carefully select the parameters of the extremum seeker \mbox{a priori} such that it converges for different unknown closed-loop dynamics without being too conservative.
Second, for the estimation of $\nabla_{\thetab}J(\thetab(k))$ in~\cite{killingsworth2006pid} a high-pass filter is necessary which reuses old output values and would require bootstrapping. Thus, a different technique must be used. Similarly, the integrator for $\thetab(k+1)$ can not have unlimited operating time. As specified below, solving these issues will require tailored modifications of the scheme from Section~\ref{subsec:extremumseeking}.

\begin{figure*}[htp!]
    \centering
    \includegraphics[trim=6.8cm 0cm 47.8cm 8.9cm, clip=true,width=\linewidth]{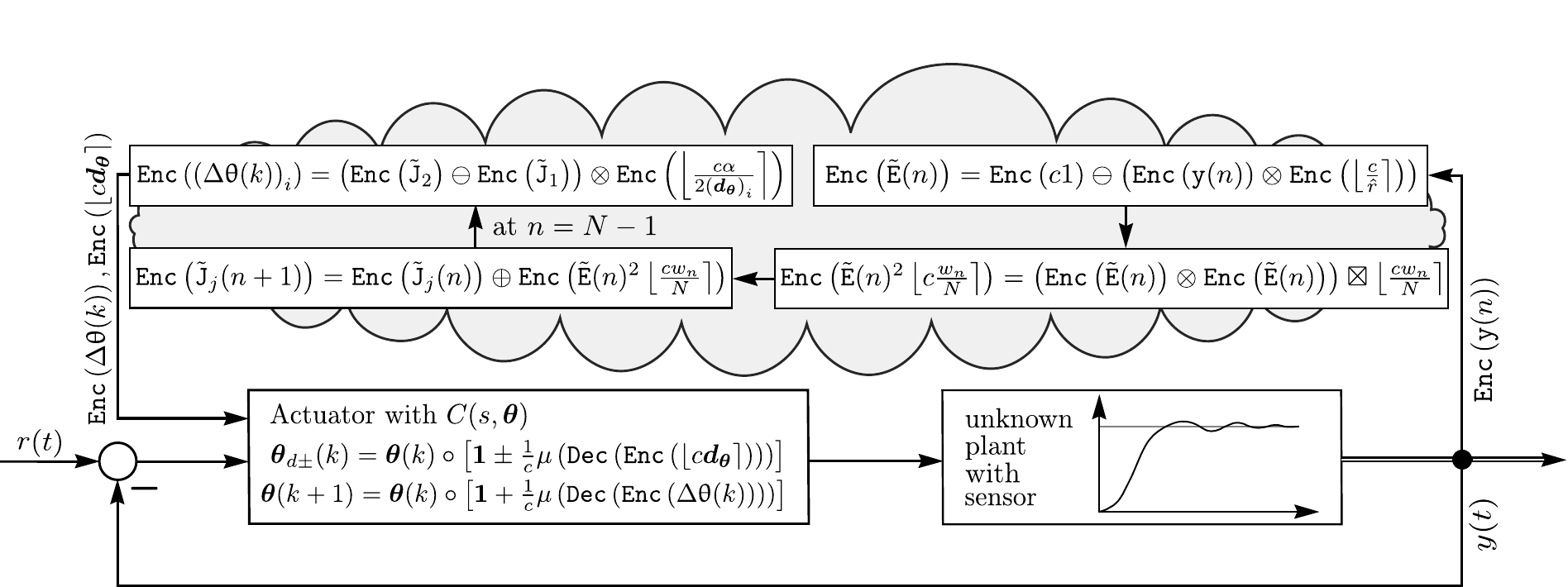}
    \caption{Encrypted cloud-based ES with scaled quantities $\yt$, $\tilde{\mathtt{E}},\tilde{\mathtt{J}}$, as well as $\text{\straighttheta}$.
    We dropped some explicit dependencies as well as the modulo-operation for brevity, and assumed ciphertext rescaling is applied after every multiplication.
    }%
    \label{fig:cloudES}%
\end{figure*}

\subsection{Robust encryptable extremum seeker for PID tuning}
\label{subsec:robust}
Before we move on, let us briefly note that PID controllers of the form~\eqref{eq:PID} are not practical in many industrial applications because high-frequency measurement noise and step-like changes in $r(t)$ can cause very large control inputs due to the derivative part. The standard solution for this issue is to introduce a first order derivative filter with the time constant $T_f$. These (more general) PID controllers of the form
\begin{align*}
    C(s,\thetab):=K_p+K_i\frac{1}{s}+K_d \frac{s}{sT_f+1}
\end{align*}
with $\thetab=\begin{pmatrix}K_p \,\,\,\, K_i \,\,\,\, K_d \,\,\,\, T_f\end{pmatrix}^\top$ will therefore be considered in the following. Due to the general framework provided by ES, this does not entail changes.

\textbf{Parameter updates.} In order to achieve a convergent ES scheme for different \mbox{a priori} unknown plants, a simple building block is as follows. Instead of absolute updates as in~\eqref{eq:perturb} and~\eqref{eq:gradientstep}, we make use of
\begin{subequations}
\label{eq:relstep}
\begin{align}
    \label{eq:relperturb}    
    \thetab_{d\pm}(k)&:=\thetab(k)\circ\left(\oneb \pm \db_{\thetab}(k)\right), \\
    \label{eq:relupdate}
    \thetab(k+1)&:=\thetab(k)\circ\left(\oneb + \Delta \thetab(k)\right),
\end{align}
\end{subequations}
where $\db_{\thetab}$ and $\Delta \thetab$ are relative updates opposed to before. This automatically takes the current size of $\left(\thetab(k)\right)_i$ into account and results in ``sensible'' updates. Consider, for example, $K_p(k)=100$ and $T_f(k)=10^{-3}$. Then, $K_p(k+1)=K_p(k)(1-0.05)$ and $T_f(k+1)=T_f(k)(1-0.05)$, i.e., a change by $5\%$, are probably useful updates, whereas $K_p(k+1)=K_p-0.05$ will likely have a negligible effect and $T_f(k+1)=T_f(k)-0.05$ results in positive feedback, which may destabilize the closed-loop. Furthermore, we move these updates to the controller. Thus, our tuning scheme will provide $\Delta \thetab(k)$ and $\db_{\thetab}$.

\textbf{Gradient estimation.} Next, we focus on the estimation of $\nabla_{\thetab} J(\thetab)$. Here, instead of using a high-pass filter in combination with a demodulation, we focus on square wave perturbation signals for $\thetab_{d\pm}$. In combination with the parameter updates, this allows for arbitrarily many tuning steps and circumvents the necessity for bootstrapping. The result of square wave perturbations is a gradient approximation based on finite differences. However, $2p$ cost function measurements are then necessary in order to approximate $\nabla_{\thetab} J(\thetab)$ which is inefficient. Instead, simultaneous perturbation (see~\cite{spall}) approximates the (relative) gradient stochastically based on
\begin{equation}
\label{eq:simul}
    \left(\nabla_{\thetab} \tilde{J}(\thetab)\right)_i := \frac{J\left(\thetab_{d+}\right)-J\left(\thetab_{d-}\right)}{2 \left(\db_{\thetab}\right)_i}
\end{equation}
with only two cost function measurements. Here, $\db_{\thetab}:=\gamma \hb$ consists of an amplitude $\gamma$ and a random perturbation vector $\hb$. The entries of $\hb$ are mutually independent zero-mean random variables, e.g., symmetrically Bernoulli distributed around $0$. Now, using~\eqref{eq:simul} results in various benefits. First, the (expected) performance is higher. More precisely, compared to other approximation techniques, simultaneous perturbations requires the least total amount of cost function evaluations until convergence~\cite{nesic2015stochapprox}. Moreover, since $\thetab_{d\pm}$ is a square wave signal, faster convergence compared to other perturbation wave forms~\cite{nesic2008choice} can be expected. Second, in comparison to~\cite{killingsworth2006pid}, the high-pass and additional perturbation signal parameters are now obsolete. Third, it is well-suited for encryption because there are $2^p$ possible values for $\db_{\thetab}$ which allows precomputing the corresponding multiplicative inverses a priori.

\textbf{Normalization.} In the current form, the magnitude of $\nabla_{\thetab} \tilde{J}(\thetab)$ may be very different for different closed-loops, which may also result in quite different $\Delta \thetab(k):=\alpha \nabla_{\thetab} \tilde{J}(\thetab)$ and destabilizing or conservative updates. In this context, it would be useful to replace $\nabla_{\thetab}\tilde{J}(\thetab)$ with the normalization $\nabla_{\thetab}\tilde{J}(\thetab)/\|\nabla_{\thetab} \tilde{J}(\thetab)\|$. Then, $\Delta \thetab(k)$ could be tuned solely by means of $\alpha$. The problem is, however, that $\|\nabla_{\thetab}\tilde{J}(\thetab)\|$ depends on $\thetab(k)$. Consequently, a costly and/or unreliable encrypted inversion would be necessary. This is why we proceed differently here. Our goal is to obtain similar magnitudes for $J(\thetab_{d\pm})$ regardless of the closed-loop response. Then, by means of~\eqref{eq:simul} with a predefined $\db_{\thetab}$, also $\nabla_{\thetab} \tilde{J}(\thetab)$ will be similar in magnitude throughout the tuning. To this end, we replace~\eqref{eq:costintegral} in~\eqref{eq:simul} by a Newton-Cotes formula
\begin{equation}
    \label{eq:discretecost}
    \tilde{J}(\thetab):=\frac{1}{N}\sum_{n=0}^{N-1} w_n \tilde{e}(n,\thetab)^2, 
\end{equation}
where $n\in\N$ denotes the $n$-th discrete time sample of $t$, $\tilde{e}(n,\thetab):=1-y(n,\thetab)/\hat{r}$ with $\hat{r}\neq0$, $w_n$ denote the integration weights, and $N$ refers to the number of samples. Note that the proposed changes amount to scaling a discrete time version of~\eqref{eq:costintegral} by $1/(T_{\mathrm{int}}\,\hat{r}^2)$, where $T_{\mathrm{int}}$ is the integration time. Thus, optimal parameter values $\thetab^{\ast}$ are not affected. Now, the normalized control error $\tilde{e}(n,\thetab)$ contributes to a decoupling of~\eqref{eq:discretecost} from $\hat{r}$. Next, in order to capture the full step response, we propose $N\approx T_{s}/\Delta t$, where $\Delta t$ is the time step and $T_{s}$ is the settling time of the closed-loop. Then, if $N$ is large enough in~\eqref{eq:discretecost}, the magnitude of $\tilde{J}(\thetab)$ becomes insensitive to variations in $T_{s}$ and $\Delta t$, which naturally happen for different closed-loops.

\textbf{Survivorship bias.} Obviously, the presented tuning approach is somewhat naive because closed-loop stability is not guaranteed during the tuning process as in other model-free tuning methods (see, e.g.,~\cite{killingsworth2006pid,lequin2003iterative}). However, as a first step towards privacy-preserving controller tuning, our adaptions of ES work surprisingly well, as we will see in Section~\ref{sec:benchmark}.

\subsection{Encrypted evaluation}
With these preparations and by means of the homomorphisms from Section~\ref{subsec:homomorph}, an encrypted evaluation of our tailored extremum seeker, in which costly computations are deliberately avoided, can be realized as follows. We first note that, once $\alpha$, $\gamma$, and $c$ are fixed, all $2^p$ encrypted perturbations $\Enc\left(\lfloor c\db_{\thetab}\rceil \modu q\right)$ and the encrypted multiplicative inverses $\Enc\left(\lfloor(c\alpha)/(2\left(\db_{\thetab}\right)_i\rceil \modu q\right)$ can be precomputed, where encrypted quantities are scaled in line with~\eqref{eq:map}. Similarly, we precompute $\Enc\left(\lfloor c/\hat{r}\rceil \modu q\right)$, $\Enc\left(\lfloor c 1\rceil \modu q\right)$, and $\Enc\left(0\right)$, where the latter serves as an initialization for the integration. During operation, $\Enc\left(\lfloor c\db_{\thetab}\rceil \modu q\right)$ is randomly selected from the preencrypted values and used as a perturbation signal. Then, after exciting the closed-loop, the cloud obtains
$\Enc\left(\yt(n)\right)$ and computes $\Enc\left(\tilde{\mathtt{E}}(n)\right)$ as well as $\Enc\left(\tilde{\mathtt{J}}_j\right)$ with a step-wise evaluation of~\eqref{eq:discretecost} for each $j\in\{1,2\}$. For the relative parameter update $\Enc\left(\Delta\text{\straighttheta}\right)$, $\Enc\left(\lfloor (c \alpha)/\left(2\left(\db_{\theta}\right)_i\right) \rceil \modu q\right)$ is selected based on the selected perturbation signal. This implementation prohibits the cloud from learning input output data or intermediate results and allows for an arbitrary amount of tuning steps.

Our encrypted scheme (with some additional details) is depicted in Figure~\ref{fig:cloudES}. There, ciphertext rescaling is applied after every homomorphic (and plaintext) multiplication. We briefly mentioned this technique before in Section~\ref{subsec:homomorph}. Without losing ourselves in technicalities, with ciphertext rescaling one can reduce the scaling factor of an encrypted plaintext  (otherwise it would be $c^6$ in $\Delta \text{\straighttheta}$ and not $c$) at the cost of also reducing the modulus (see~\cite{brakerski2014leveled,Cheon2017_CKKS} for details). Thus, the overflow problematic is not resolved. However, apart from some advantages, it is also convenient to use and often automatically performed in popular libraries. An implementation according to Figure~\ref{fig:cloudES} can, e.g., be realized with the cryptosystems from \cite{brakerski2014leveled,Cheon2017_CKKS}.

Finally, the remaining parameters are specified as follows. We assume that the number of tuning steps $k_{\max}$, $\hat{r}$, and $\tilde{T}_{s}$ are user-defined, where $\tilde{T}_s$ is an estimation of $T_s$. We further note that, $N$ as well as $\Delta t$ are public, since an attacker can always count the number of $\Enc(\yt(n))$ and measure the time between $\Enc(\yt(n+1))$ and $\Enc(\yt(n))$. Consequently, also $\tilde{T}_s=N\Delta t$ and $y(N)\approx y(N+l)$ for some $l\in\N\setminus\{0\}$ are public.\footnote{In fact, this enables an encrypted estimation of $\tilde{T}_{s}$ based on comparisons of $\Enc(\yt(n))$.}

\section{Numerical benchmark}
\label{sec:benchmark}
For comparison, we use the examples from~\cite{killingsworth2006pid} and investigate the plants with normalized amplification
\begin{align*}
    \{G_j(s)\}_{j=1}^{3}\!=\!\left\{\frac{\mathrm{e}^{-5s}}{20s+1},\frac{1}{\left(0.01 s+1\right)^8},\frac{1-5s}{200 s^2+30s+1}\right\}\!,
\end{align*}
where $G_1$ has a time delay, which is replaced with a third order Padé approximant, $G_2$ is a high order model with repeated poles, and $G_3$ is non-minimum phase. Note that we significantly decreased the time constant of $G_2$ in comparison to~\cite{killingsworth2006pid} in order to test our scheme with a greater variety of dynamics. Next, $\{\Delta t_j\}_{j=1}^{3}\!\!=\!\{0.01,10^{-4},0.01\}$ and $\{\tilde{T}_{s,j}\}_{j=1}^{3}\!\!=\!\{50,0.05,80\}$ are used for simulation, where initial controller parameters $\thetab(0)$ stem from the Ziegler-Nichols oscillation method. We fix $\alpha=1$, $\gamma=10^{-2}$, and $k_{\max}=50$ for all tunings and use a trapezoidal rule 
in~\eqref{eq:discretecost}. For the cryptographic implementation we opted for the leveled homomorphic RNS Variant of CKKS~\cite{Cheon2017_CKKS} available in the PALISADE library~\cite{palisade} which is executed on an Intel i7-7600U. In this context, $q\approx 2^{160}$ with $4$ levels, and a ring dimension of $2048$ provide a parameter-dependent security of approximately $80$ bits according to the estimator\footnote{\href{https://bitbucket.org/malb/lwe-estimator}{https://bitbucket.org/malb/lwe-estimator}} from~\cite{albrecht2015concrete}.

\begin{figure*}[htp!]
    \centering
    \subfloat{{\includegraphics[trim=3cm 8.5cm 2.5cm 8cm, clip=true,width=.33\linewidth]{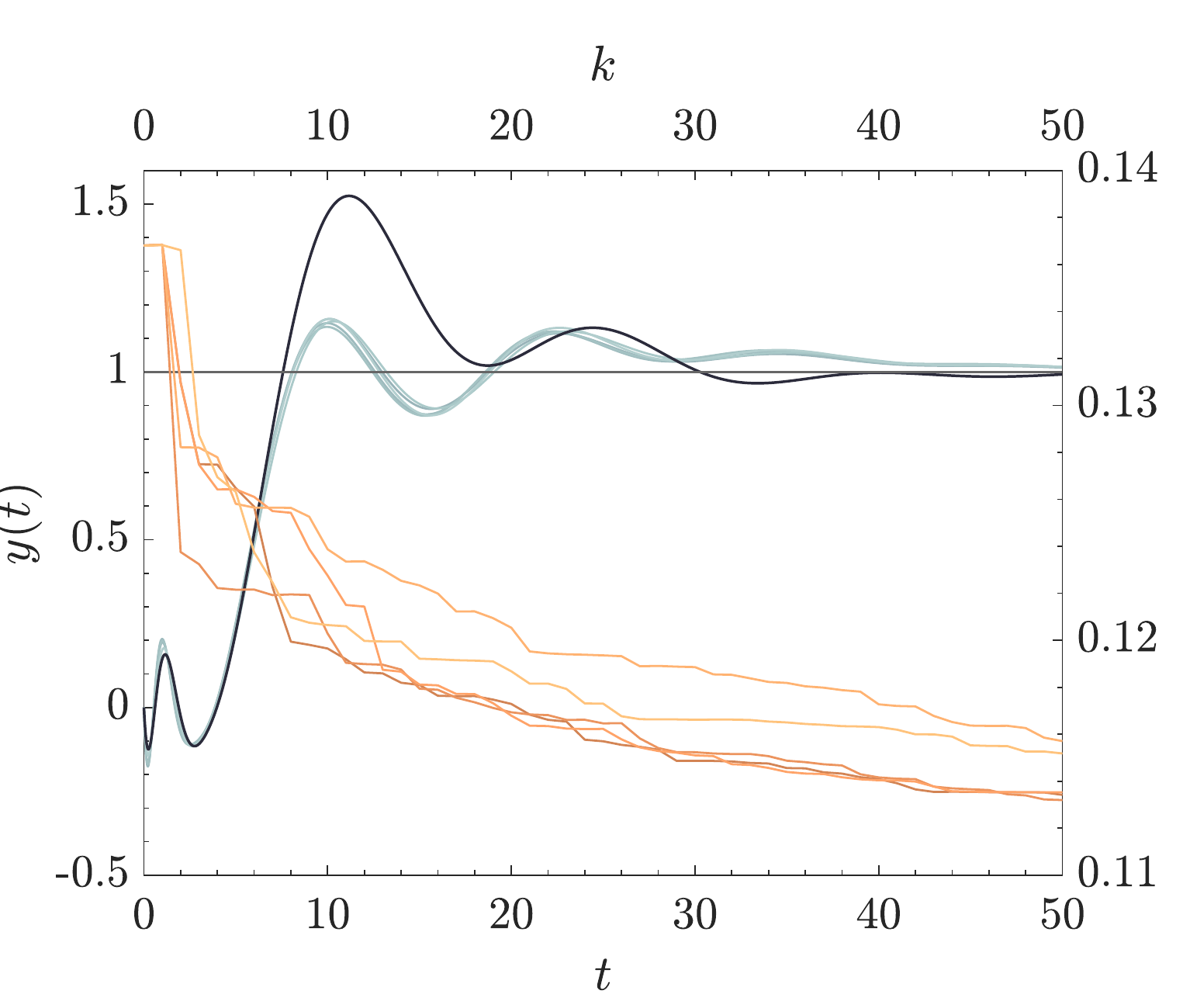} }}%
    \subfloat{{\includegraphics[trim=3cm 8.5cm 2.5cm 8cm, clip=true,width=.33\linewidth]{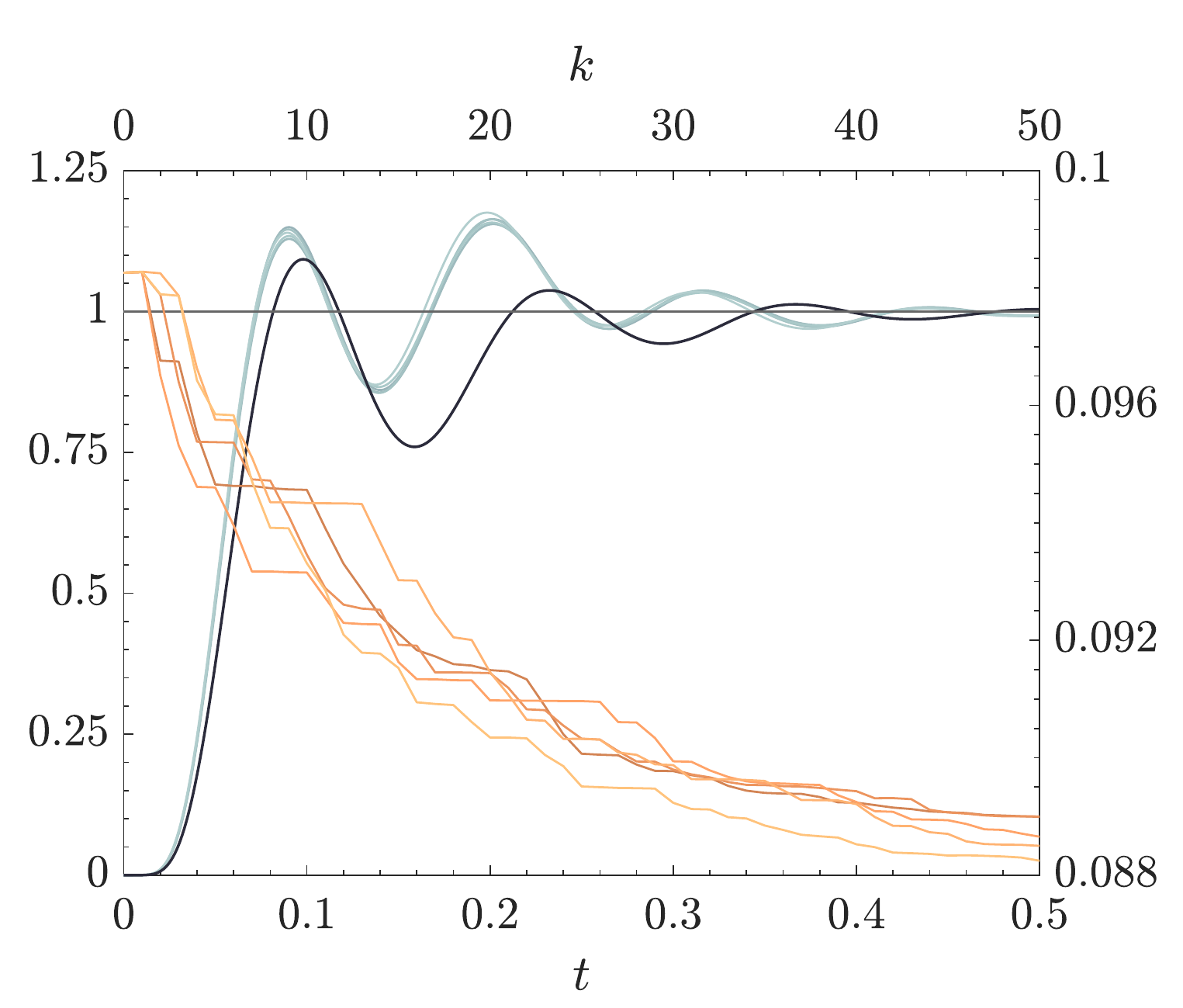} }}%
    \subfloat{{\includegraphics[trim=3cm 8.5cm 2.5cm 8cm, clip=true,width=.33\linewidth]{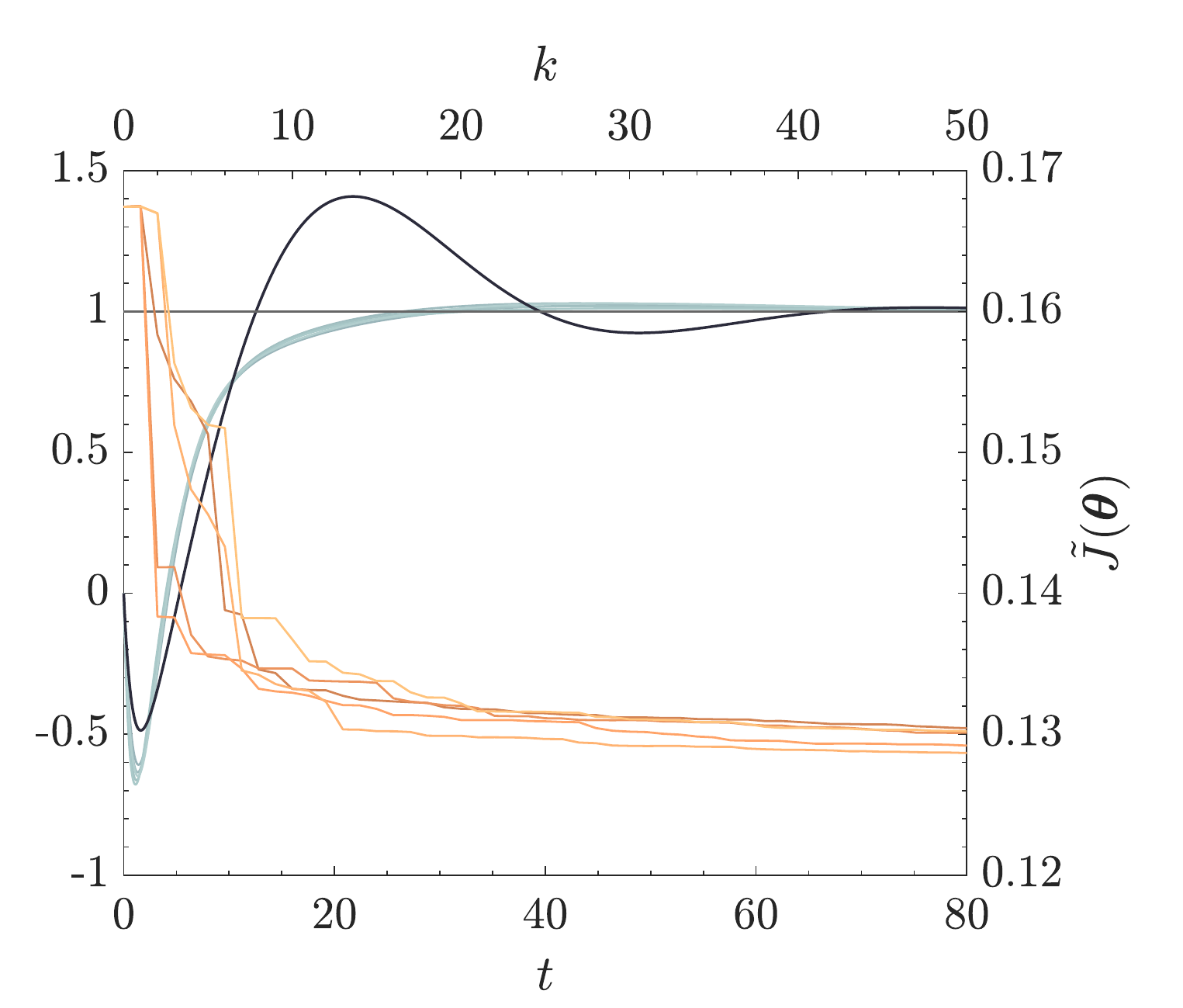} }}%
    \caption{Results of five encrypted ES-based PID tunings. Initial and final closed-loop responses $y(t)$ for $G_1(s),G_2(s)$, and $G_3(s)$ in black and light-gray, respectively. The corresponding cost functions $\tilde{J}(\thetab)$ are depicted in orange over $k$.}%
    \label{fig:different plants}%
\end{figure*}

\textbf{Tuning results}. We start with the tuning results of our proposed scheme for $G_1$, $G_2$, and $G_3$ in Figure~\ref{fig:different plants}.
Here, the control errors in all closed-loops are significantly reduced, in spite of the different dynamics. Furthermore, we present five tunings for each plant to take the stochastic effects of~\eqref{eq:simul} into account. Clearly, all tunings converged with similar speed and result in almost identical performance. Next, the parameters corresponding to Figure~\ref{fig:different plants} can be found in Table~\ref{tab:results}.
\begin{table}[t]
\caption{Tuning results for $G_1$(s), $G_2(s)$, and $G_3(s)$.}
\label{tab:results}
\centering
\begin{threeparttable}
\begin{tabular}{c cc cccc}
\toprule
&$k$ &$\sigma^{\dagger}$ & $K_p$ & $K_i$ & $K_d$ & $T_f$ \\
\midrule
\multirow{3}{*}{$G_1(s)$} 
& $0$                 & --  & $4.08$ & $0.45$ &  $9.33$ & $0.50$ \\ 
& \multirow{2}{*}{50} & $0$ & $3.24$ & $0.22$ &  $9.93$ & $0.36$ \\ 
&                     & $5$ & $3.64$ & $0.22$ & $10.33$ & $0.10$  \\
\midrule
\multirow{3}{*}{$G_2(s)$} 
& $0$                 & --  & $1.11$ & $14.61$ & $0.02$ & $1\times10^{-3}$ \\ 
& \multirow{2}{*}{50} & $0$ & $0.96$ & $22.94$ & $0.03$ & $9\times10^{-4}$ \\ 
&                     & $5$ & $0.81$ & $22.15$ & $0.04$ & $7\times10^{-4}$ \\ 
\midrule
\multirow{3}{*}{$G_3(s)$} 
& $0$                 & --  & $3.53$ & $0.21$ & $14.82$ & $0.50$ \\ 
& \multirow{2}{*}{50} & $0$ & $2.72$ & $0.10$ & $19.83$ & $0.40$ \\ 
&                     & $5$ & $3.02$ & $0.09$ & $18.25$ & $0.39$ \\ 
\bottomrule
\end{tabular}
\begin{tablenotes}
    \item[$^{\dagger}$ variance of the measurement noise in $\%$ of $y_{\infty}$] 
    \end{tablenotes}
\end{threeparttable}
\end{table}
There, also Gaussian measurement noise is considered. Although a significant amount of noise is used for each plant, we can see that the parameters are very similar compared to the noise-free case and none of the tunings led to an unstable closed-loop, which is a result of the integration~\eqref{eq:discretecost}. Interestingly, more iterations reveal that the convergence is slower in the presence of noise. 

Now, let us focus on the effect of $\tilde{T}_s$ and $\Delta t$. With a view on~\eqref{eq:discretecost} it becomes clear that both of these parameters influence $\tilde{J}(\thetab)$ by means of $N=\tilde{T}_s/\Delta t$. On one hand, $N$ could be much larger than necessary, i.e., $\tilde{e}(n,\thetab)\approx0$ for many $n$ in~\eqref{eq:discretecost}. Then, $\tilde{J}(\thetab)$ will be smaller in comparison to the case where $N$ is adequately chosen, which reduces the magnitude of $\nabla_{\thetab}\tilde{J}(\thetab)$ and, as a result, the convergence speed. On the other hand, if $N$ is much smaller than necessary, $\nabla_{\thetab}\tilde{J}(\thetab)$ might become larger than expected, which can cause a non-stable update for $\thetab$. However, changes of $N$ had negligible effects on the tuning outcomes in our studies. For example, unstable behavior occurs the fastest for $G_2$ if $N$ is reduced by $70\%$.

Lastly, the cryptographic performance is as follows. The encryption of $\yt(n)$ and the decryption of $\Enc\left(\Delta \text{\straighttheta}(k)\right)$ as well as $\Enc\left(\lfloor c\db_{\thetab}\rceil \modu q\right)$ take approximately $2\,\mathrm{ms}$. By means of~\eqref{eq:discretecost} the homomorphic evaluation depends on $N$. In particular, for $N=1$ an evaluation of $\Enc\left(\Delta \text{\straighttheta}(k)\right)$ requires around $11\, \mathrm{ms}$. The values for $N$ we used in the examples therefore lead to evaluation times of in the order of $5.5\, \mathrm{s}$ to $55\,\mathrm{s}$. Hence, faster evaluations require smaller values for $N$ at the cost of less robustness.

\section{Conclusion and Outlook}
\label{sec:outlook}
As a first step towards privacy-preserving controller tuning as-a-Service, we tailored an extremum seeking algorithm for encrypted PID tuning. More precisely, relative parameter updates, a stochastic gradient approximation, and a normalized objective function provided a more ``encryption-friendly'' reformulation, while contributing to the robustness (and performance) of our algorithm. We tested our encrypted extremum seeker against different plant dynamics, measurement noise, as well as parameter variations. In all cases, the PID parameters were significantly improved, which shows the effectiveness of our approach.

We are well aware of the fact, that the simplicity of the algorithm makes outsourcing it to a cloud questionable. Apart from that also stability during the tuning process is an issue. Therefore, one may view the proposed method as a first step towards CaaS.
For future research, the design and encryption of other tuning algorithms will be of interest. With a view on the rapid performance enhancements of HE over the last decade, it may soon be possible to evaluate more complex controller designs in an acceptable time.

\section{Acknowledgment}
The authors gratefully acknowledge the support by the German Research Foundation (DFG) under the grant SCHU 2940/4-1.

% References

\end{document}